# Total Cross Sections

Giorgio Giacomelli

Dipartimento di Fisica dell'Universita' di Bologna, and INFN, I-40127 Bologna, Italy

Giacomelli@bo.infn.it

**Kycia Memorial Symposium**

Brookhaven National Laboratory, Upton, N.Y. 11973-5000, U.S.A.

Friday May 19, 2000

## 1. Introduction

The measurements of the hadron-hadron total cross sections are the first measurements performed when a new hadron accelerator opens up a new energy region; the measurements were made as function of the incoming beam momentum or c.m. energy and have often been repeated with improved accuracy and finer energy spacing.

Most of the systematic total cross section measurements of the long-lived charged hadrons ($\pi^\pm$, $k^\pm$, $p^\pm$) on hydrogen and deuteron targets at fixed target accelerators were performed using the transmission method pioneered at Brookhaven National Laboratory; the method is capable of high precisions, typically point to point precisions of 0.1 - 0.2 % and a systematic scale uncertainty of 0.4 – 1.0 %.

Fig. 1, from the Data Particle Group, shows the behaviour with energy of the total cross sections of different hadrons. At low energies, in the so called *resonance region*, one observes a number of peaks and structures which decrease in size as the energy increases. Above 5 GeV lab momentum, in the *continuum region*, there are no more structures: the cross sections decrease smoothly, reach a minimum and then slowly rise with increasing energy (the *asymptotic region*). In the low energy region the cross sections depend strongly on the type of colliding hadrons and on the total isotopic spin, whilst in the high energy region these dependences tend to disappear as the energy increases.

I shall recall some of the measurements and of the discoveries made by Ted Kycia and by our colleagues, some of whom are present in the audience. Besides Ted, we are missing Rod Cool, who was the leader and a driving force for the measurements.

Among Ted's papers, I found his Curriculum Vitae, probably written around 1974; it is written in a very simple form and it well states Ted's interests and achievements in total cross section measurements; the part concerning total cross section measurements is reproduced below.

---

Curriculum Vitae

## THADDEUS F. KYCIA

... Since 1960 he has led the development and construction of differential Cherenkov counters for use in charged secondary beams. Their function has been to electronically identify selected types of particles with a very high rejection of all other particles. The counters which were built spanned the full range of velocities available at the AGS. Most recently, he designed and built two gas differential Cherenkov counters for use at Fermi National Laboratory in the total cross section experiment. The resolution of one counter was adequate to separate K mesons from mesons even at a momentum of 200 GeV/c.

Over the last decade Ted Kycia led an effort to improve the precision with which total cross sections of charged particles could be measured. This led to the dicovery of six massive meson-nucleon resonances. This was followed by a series of measurements of total cross sections of $K^-$ mesons, $K^+$ mesons and antiprotons on nucleons in search of structure. The high precision systematic measurements revealed a large number of previously unobserved resonances and structures. A number of hyperon resonances were found in both isotopic spin zero and isotopic spin one states. A number of structures were found in the antiproton-proton total cross section which would be due to previously unobserved massive pion resonances. In the $K^+$ meson-nucleon total cross section measurements a number of structures also were discovered. These have subsequently been studied by groups in this country and in Europe. The question of whether any of the $K^+$ meson-nucleon structures could be due to the existence of exotic $Z^*$ 's is still unresolved.

The techniques developed at BNL for measuring total cross sections to a high precision were then applied to the measurement of total cross sections at FNAL. This latter experiment was carried out in collaboration with physicists from Rockefeller and Fermilab. ...

---

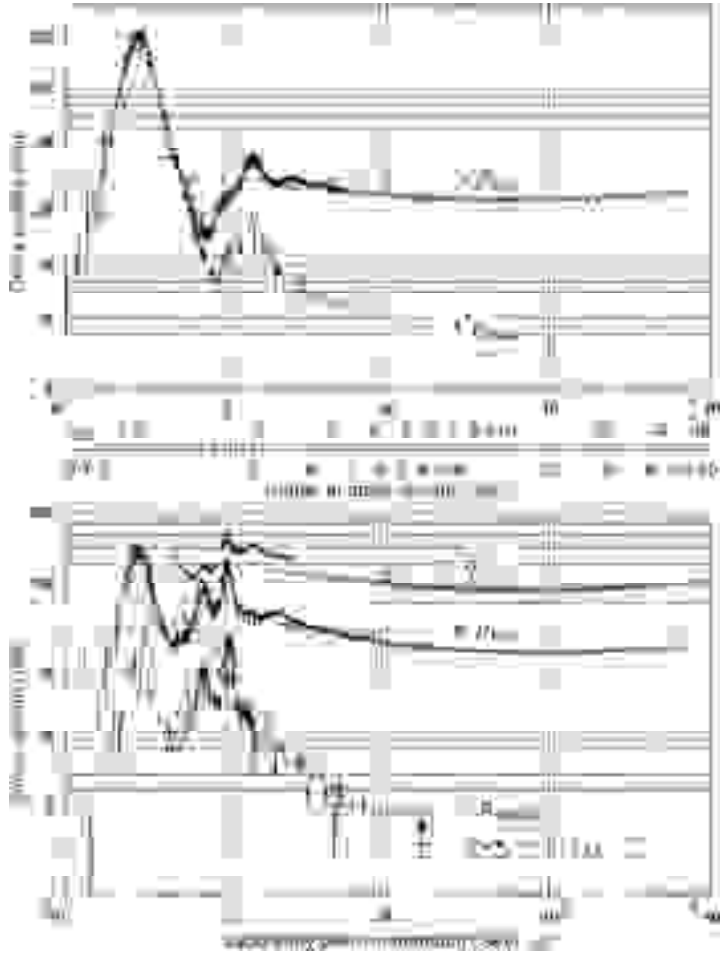

Fig. 1. Compilations of data of the total and of the integrated elastic cross sections versus lab momentum for $\pi^{\pm}p$ and $\pi^{\pm}d$ scattering [Particle Data Group].

## 2. Total cross sections at Brookhaven

At BNL a series of measurements were made with different beams covering the resonance region [1-2, 4-9] and the beginning of the continuum region [3]. The cross sections were measured on protons and deuterons and also on various nuclei [9].

A systematic and precise total cross section measurement in the resonance region may be thought of as a method for detecting the existence of new resonances: this was the main aim of the Brookhaven measurements. Low mass rersonances are easy to detect because they produce large effects. Higher mass resonances, however, show up as broad and non prominent structures, often overlapping with one another, so that one needs to measure the total cross sections with high precision at many closely spaced points. Errors in the absolute values can be tolerated only if they are essentially energy independent.

Ted had a particular way of searching by eye for new mini-structures at relatively high energies: he carefully plotted the measured points on a large graph, which he then was viewing along the points, trying to see if there were mini-structures.

The $\pi^+p$, $K^+p$, and pp are pure isospin states. In the other cases one has a mixture of two isospin states. The determination of the pure isotopic spin cross sections requires the measurement of two cross sections, which involves changing either the incident or the target particle. For pions it is easy to measure both $\pi^+p$ and $\pi^-p$ total cross sections, and hence to derive the total cross sections $\sigma_{1/2}$ and $\sigma_{3/2}$ for pure isospin states. For the other cases the simplest solution is to measure the cross sections off protons and off neutrons. Unfortunately, the best neutron target is a bound neutron-proton state (the deuteron), so that in this case the problems of nuclear physics in the deuteron severely limit the analysis of the data. A careful unfolding procedure has to be performed to extract the pure isospin cross sections.

If a structure is found in a total cross section measurement, and assuming that it is due to a resonance, the information that it yields includes the mass $M$ (the c.m. energy of the peak), the width at half height $\Gamma$, the height $\sigma_R$, and the isotopic spin $I$.

The total cross section method of resonance-hunting must be considered as a "coarse spectrometer", with a not-so-good energy resolution. In fact "spectrometers" with higher resolutions have been used, like elastic scattering and "phase-shift analyses".

Total cross section measurements do not provide enough information to establish conclusively that a peak in a definite isospin state is a resonant state, i.e. a state with definite quantum numbers. In fact, a structure could also come from a threshold effect, such as the opening up of a new important channel, or other kinematical effects.

The principle of the method used for measuring total cross sections is that of a standard transmission good geometry experiment.

The low energy beams were partially separated secondary beams, see for example Fig. 2.

After momentum and mass separation, the beam is defined by a sysytem of scintillation counters and by a Cherenkov counter, which further electronically distinguishes between wanted and unwanted particles. The beam then alternatively passes through a hydrogen, deuterium, or dummy target and converges to a focus at the location of the transmission counters, each of which subtends a different solid angle at the center of the target (see Fig. 3). The method is thus to evaluate the partial cross sections $\sigma_i$ measured by each individual transmission counter and to extrapolate these cross sections to zero solid angle to obtain the total cross section.

The momentum spread of a beam was typically 0.75 %. With a circulating beam of $10^{12}$ protons the used $K^+$-meson flux was about 3500 per pulse at 1.6 GeV/c, increasing at lower energies. The $K^-$ meson fluxes were typically one third of the $K^+$-meson fluxes. The antiproton fluxes reached about 10000 $\bar{p}$'s per pulse at 2.5-3.0 GeV/c, and smaller values at lower momenta.

The separation achieved with the electrostatic separators was considerable, giving at worst a ratio of 2:1 between the wanted and unwanted particles; the contamination of unwanted particles was assured to be less than 0.1 % by the Cherenkov counters.

Figures 4 to 8 show some of the experimental results obtained. Figs. 4 and 5 show the K⁻p, data and the K⁻N, I=0 and I=1 data, respectively. Notice the quality of the data in Fig. 4 and Fig. 7 compared to previous measurements. Fig. 5 shows the I=0 and the I=1 total cross sections for the K⁻N scattering after proper unfolding. Notice the structures which decrease in size as the energy increases. For small structures at the highest energies, global analyses do not still give final conclusions on their parameters.

The pion nucleon system, Fig. 6, is clearly overdetermined since one can measure the $\pi^-$p, $\pi^+$p, $\pi^-$d, $\pi^+$d total cross sections.

In the $\bar{p}p$ system the structures in the covered energy range are very small (Fig. 7).

In the K⁺N, I=0 state there is a structure at the center of mass (c.m.) energy of about 1910 MeV (Fig. 8). Many measurements have been made on this system, without reaching a final conclusion, though a possible I=0 resonant state seems to be indicated for this "exotic system" [10]. Further work on this system may be worthwhile.

During the period of total cross section measurements at BNL other important measurements were made there: (i) the first measurements of the magnetic moments of the hyperons, (ii) elastic scattering measurements with the discovery of the shrinking of the diffraction pattern, (iii) bubble chamber measurements with the discovery of the Ω⁻ [11, 12]. The scientific atmosphere at BNL was at its best. But also the human atmosphere was at its best; in particular the collaborators in the total cross section measurements became friends and the frienship lasted for all subsequent measurements at other accelerators.

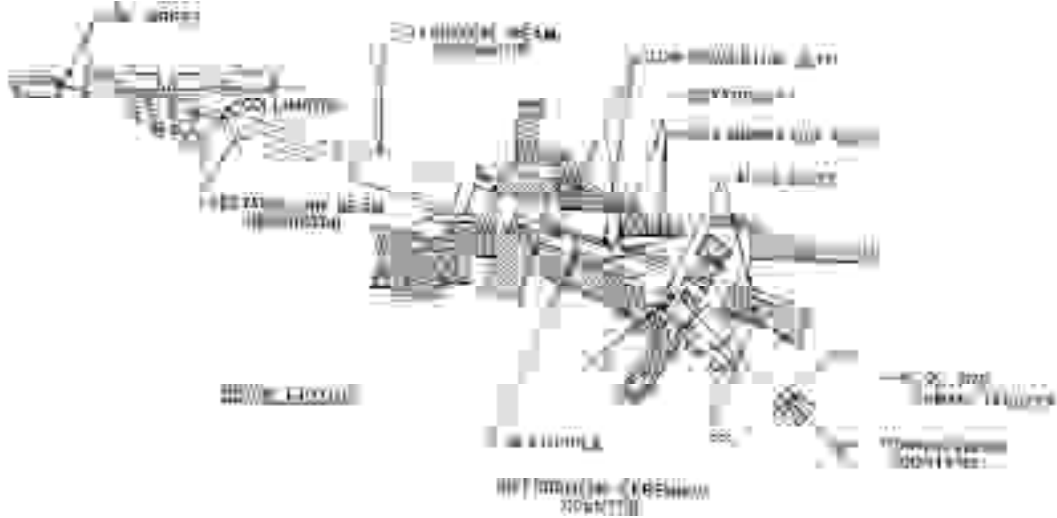

Fig. 2. Layout of a partially separated Brookhaven beam. $Q_1$-$Q_8$ are quadrupoles; $D_1$-$D_8$ are bending magnets; S1-S3 and $\bar{G}$ are scintillation counters. Note the electrostatic beam separators.

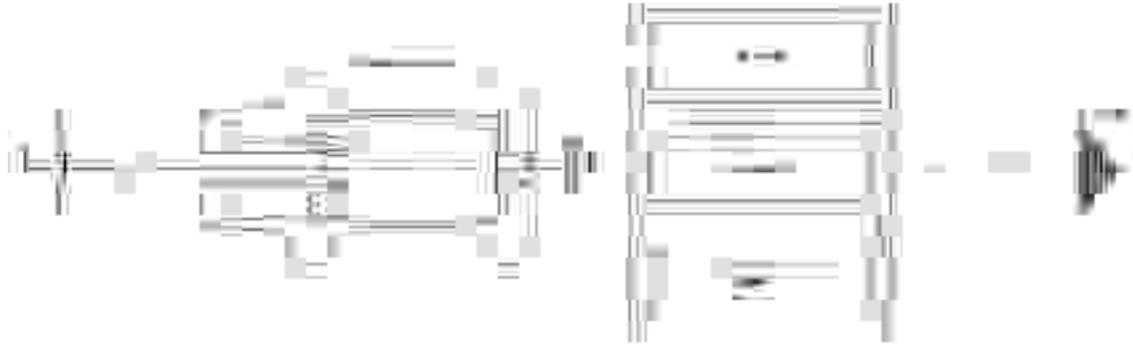

Fig. 3. Layout of the experimental apparatus for the measurement of low energy total cross sections. $S_1$, $S_2$, $S_3$, and $\overline{G}$ are scintillation counters defining the beam. The gas differential Cherenkov counter is shown; it was replaced by a liquid differential Cherenkov counter for measurements at lower momenta. $H_2$, $D_2$, and the dummy are the liquid hydrogen, liquid deuterium and the dummy targets, respectively. $S_4$-$S_{12}$ are the transmission counters.

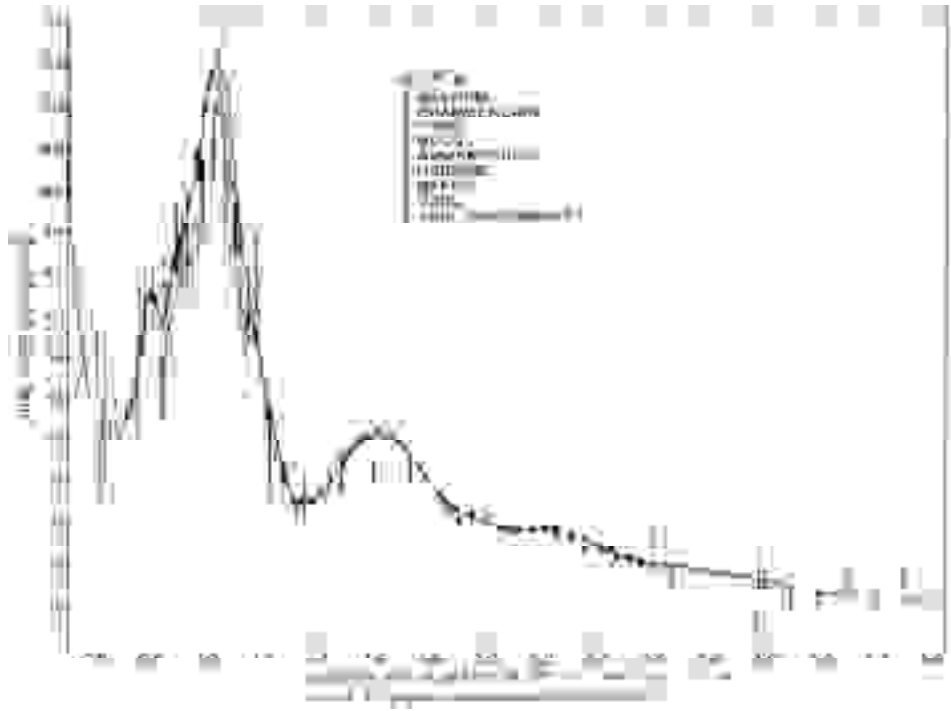

Fig. 4. K⁻-p total cross sections in the range 0.6-3.5 GeV/c lab. momentum.

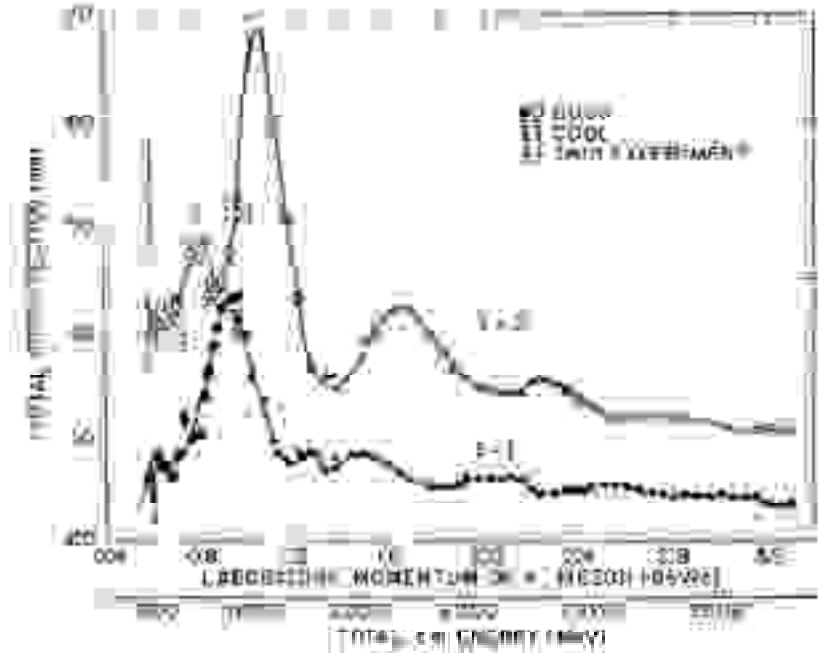

Fig. 5. K⁻N total cross sections in the pure I=1 and I=0 states, for the range 0.6-3.3 GeV/c.

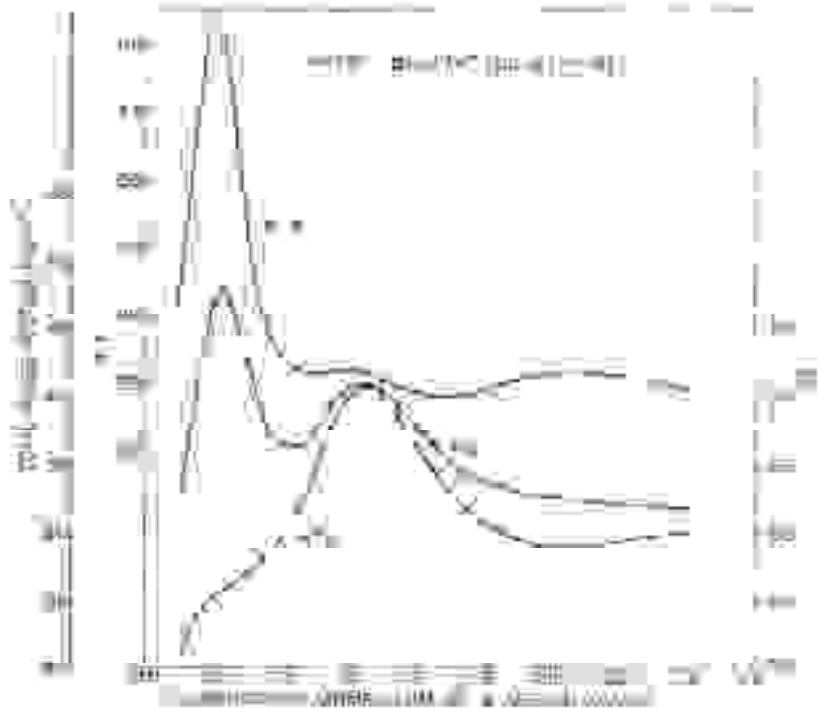

Fig. 6. ⁻p, ⁺p, and ±d total cross sections. For all points shown, the statistical errors are smaller than the size of the data points.

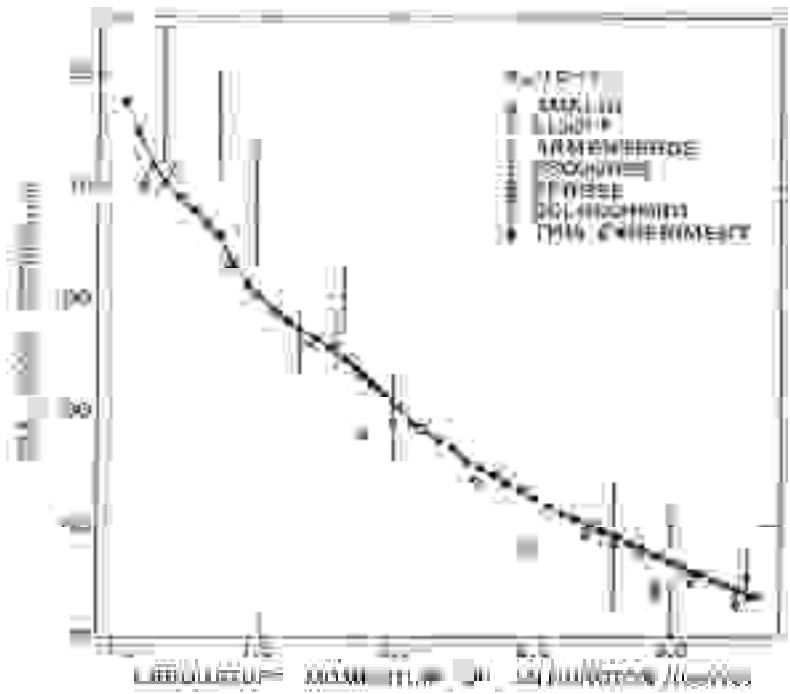

Fig. 7. $\bar{p}p$ total cross sections, in the range 1.0-3.3 GeV/c. For each point, the statistical error is less than the size of the dot.

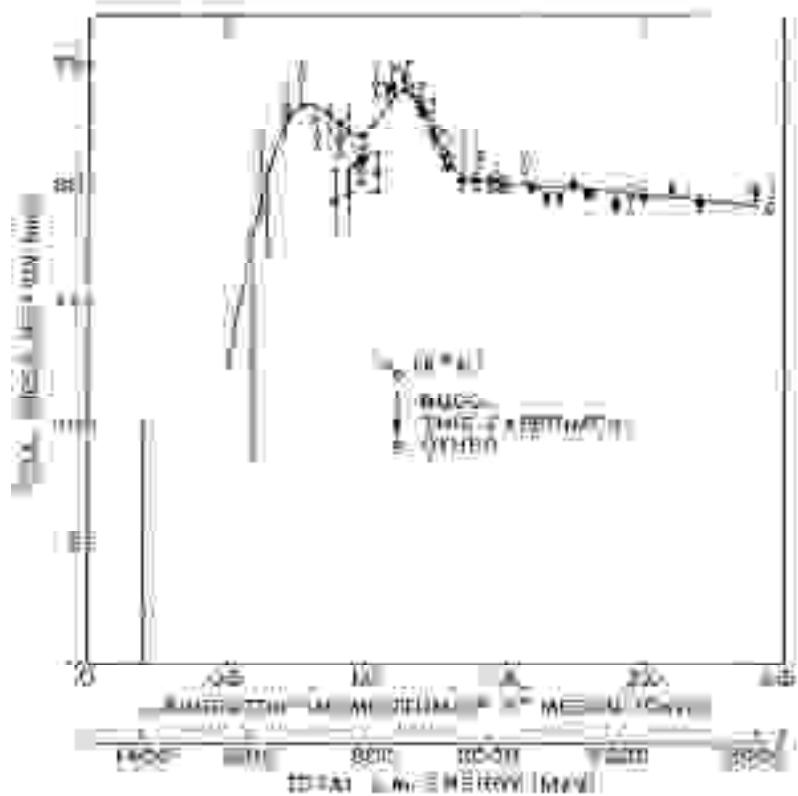

Fig. 8. Total cross section $\sigma_0$ for the I=0 isotopic spin state for the $K^+N$ system.

## 3. Total cross sections at Fermilab

Total cross section measurements at intermediate energies have been performed at BNL [3] and at Serpukhov [13-15]. Then followed measurements at the CERN-ISR at high energies [15, 16] and two sets of measurements at the Fermilab fixed target accelerator [17-22].

Fig. 9 shows the layout of the total cross section measurements at Fermilab. The differences compared to the Brookhaven measurements were mainly due to the much higher energies of the Fermilab beams, thus to the impossibility of using electrostatic separators, to the need of much more selective differential Cherenkov counters, longer targets, etc. Incident particles were defined by scintillation counters and identified by two differential gas Cherenkov counters, allowing cross sections of two different particles to be measured simultaneously. In addition, a threshold gas Cherenkov counter could be used in anticoincidence when required. Typical particle separations are shown in Fig. 10 using two counters made by Ted. Sufficient $\pi^+$-$K^+$ separation was achieved up to 340 GeV/c and at higher momenta using corrected optics [12] Contamination of unwanted particles in the selected beam particles was always below 0.1 % thanks to the marvellous Cherenkov counters designed and built by Ted. In the pion and kaon beams there were small admixtures of muons and electrons (at the level of one part in a thousand and 1 %, respectively). Electrons in the gas Cherenkov counter pion signal were identified by their characteristic signal in a 22-radiation length lead-glass Cherenkov counter placed downstream of the transmission counters. Muons were identified by their ability to pass through 5 m of steel placed downstream of the transmission counters. Other differences concerned the order in the transmission counters (first the large transmission counters at Brookhaven, and the reverse at Fermilab: this is one problem which usually leads to strong debates inside a group!). The transmission counters could be moved on rails so as to subtend at each energy the same t-range. The targets were 3 m long, much longer than in the BNL experiments. The three targets ( hydrogen, deuterium and dummy ) were surrounded by a common outer jacket of liquid hydrogen for temperature stability. The vapour pressure was continuously monitored and the hydrogen and deuterium densities were determined; their density variations were less than 0.07 %. Target lengths were measured to 0.03 %. Repeated measurements indicated that the cross section measurements were stable to better than 0.2 %.

The data were taken first in the range 50 to 200 GeV/c secondary beam momentum, and later in the range 200 to 370 GeV/c. A compilation of the measured data is presented in Fig. 11. These new higher energy data were presented at the 1974 High Energy International Conference in London. In this occasion a press conference was made and the statements presented there summarize the interest of the measurements and their possible interpretation.

Press Conference

[...]

BROOKHAVEN, FERMILAB AND ROCKEFELLER UNIVERSITY PHYSICISTS OPEN NEW WINDOW ON STRONG NUCLEAR HIGH ENERGY INTERACTIONS

London, July 2, 1974: a new experiment announced today at the 17$^{th}$ International Conference on High Energy Physics held at Imperial College in London, England indicates the measurements. The solid line is a fit of the $\sigma_{tot}$ and $\rho$ data with dispersion relations; the region of uncertainty is delimited by dashed lines surprisingly systematic character of the interaction of the fundamental particles of matter at high energies.

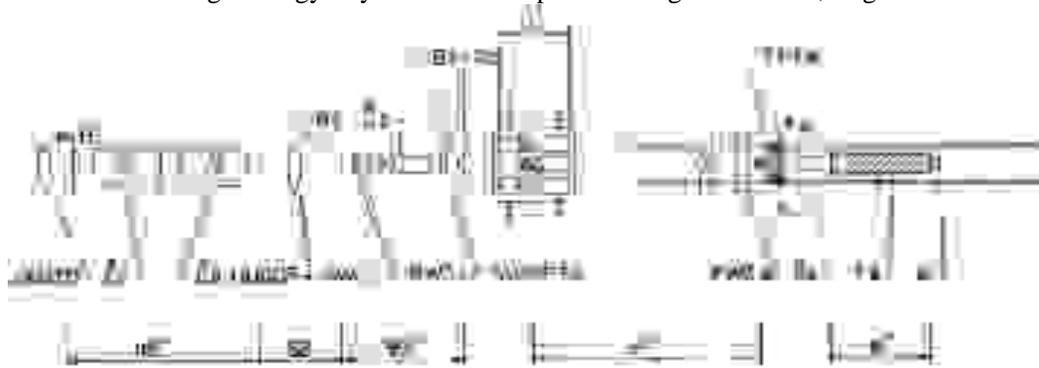

Fig. 9. Layout of the experimental apparatus for the measurement of the high energy total cross sections at Fermilab. $C_1$, $C_2$ ($C_3$) are gas differential (threshold) Cherenkov counters, PWC1-PWC3 are proportional wire chambers $B_1$-$B_3$ and $A_1$-$A_2$ are scintillation counters. $H_2$, $D_2$, VAC are the liquid hydrogen liquid deuterium and dummy targets, $T_1$-$T_{12}$ are the transmission counters, $_1$-$_2$ are scintillation counters used for efficiency measurements, $C_e$ is a lead glass Cherenkov counter; the iron absorber and the muon (μ) scintillation counter were used for estimating the muon contamination.

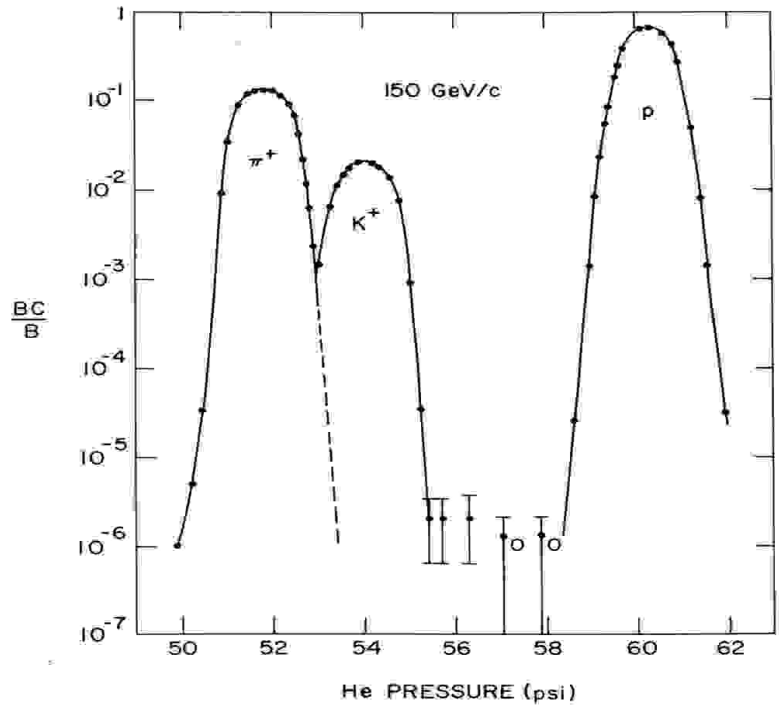

Fig. 10. Relative counting rate versus helium pressure in the gas Cherenkov counters for the beam of unseparated particles of 150 GeV/c momentum.

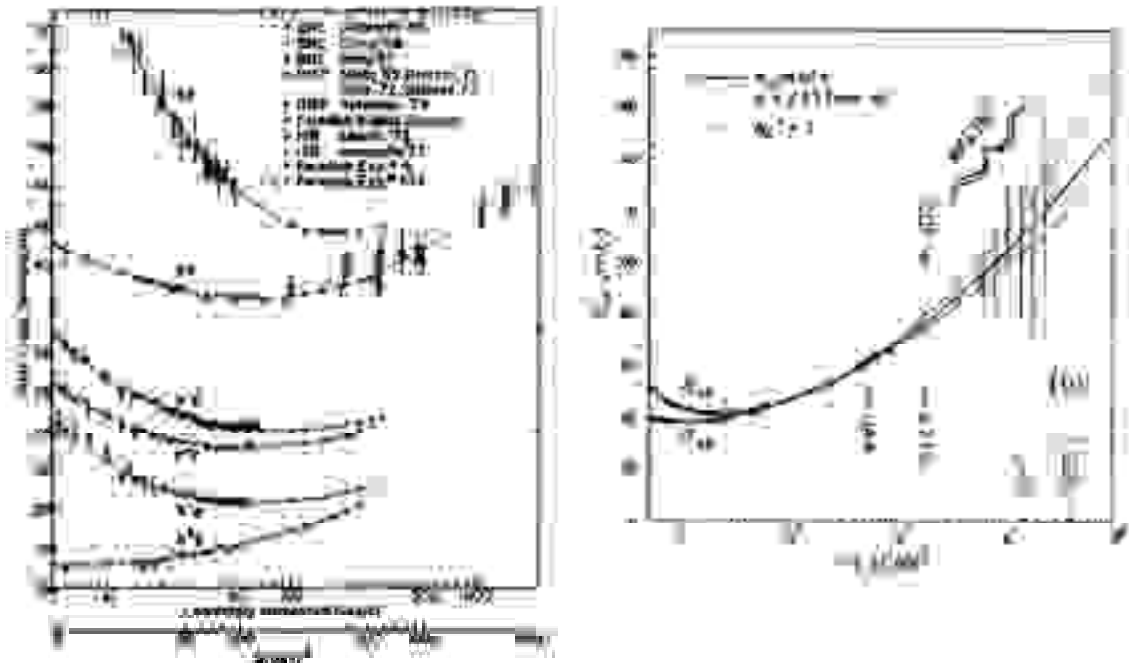

Fig. 11. (a) Compilation of $\bar{p}p$, $pp$, $\pi^-p$, $\pi^+p$, $K^-p$ and $K^+p$ total cross sections plotted versus c.m. energy. (b) the $\bar{p}p$ and the $pp$ total cross sections, including cosmic ray

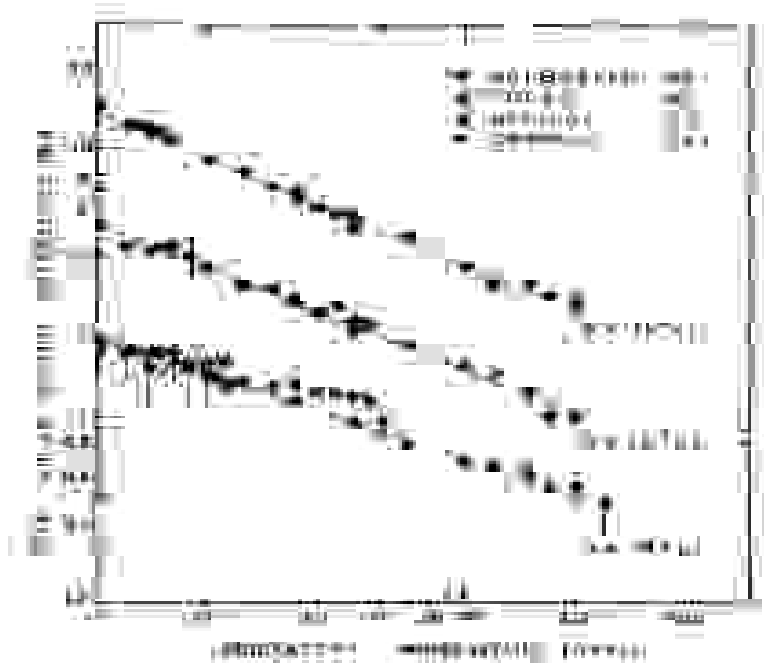

Fig.12. The differences of total cross sections for $\pi^\pm$, $K^\pm$, $p$ and $\bar{p}$ interactions with protons. The solid lines represent fits of the data to a power law dependence.

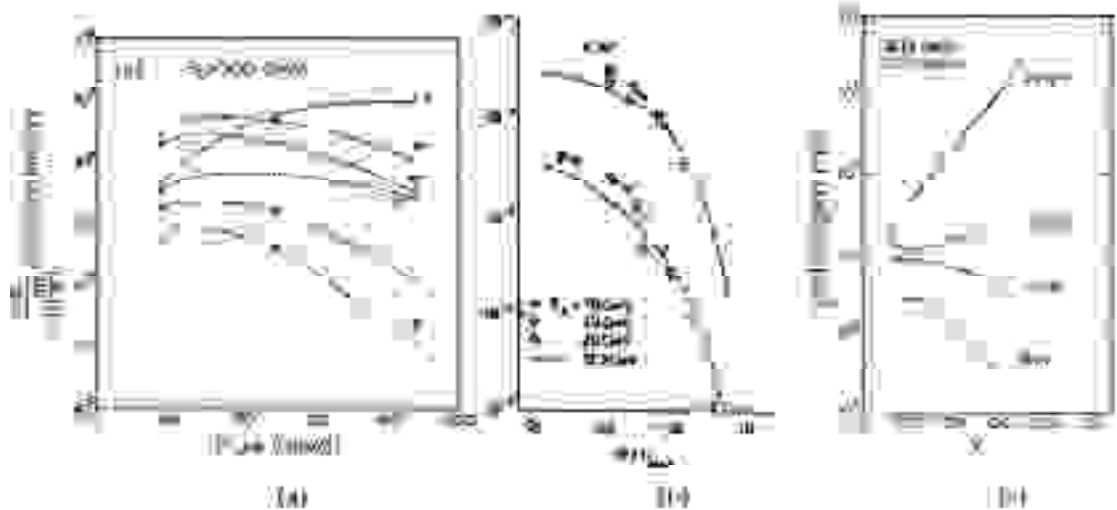

Fig.13. (a) Production cross sections of the six long-lived charged hadrons plotted vs lab momentum; (b) (c) particle ratios vs $p/p_{max}$.

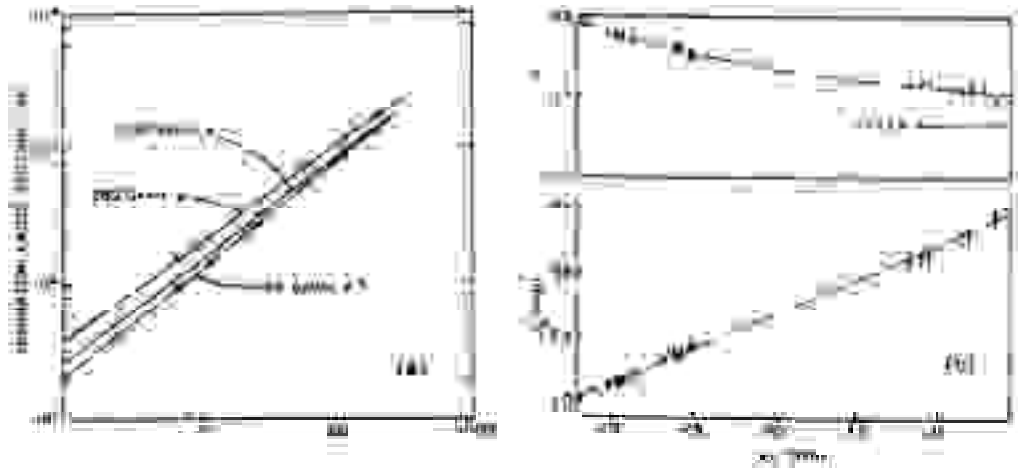

Fig. 14. Absorption cross sections in different nuclei.

The report describes the results of bombarding protons and neutrons, the basic constituents of atomic nuclei, with six different types of very high energy sub-nuclear particles produced by the new U.S. Fermi National Accelerator Laboratory located near Chicago, Illinois.

The Measurements allow a precise comparison of the interaction probabilities of each of the six different strongly interacting probe particles with the proton and neutron. These interaction probabilitiers are usually referred to as effective areas or "total cross sections" of the proton and neutron.

These precise measurements, with an accuracy of about one part in 500, reveal that the effective size of both the proton and neutron increase for five of the six probes when their energy is increased from 50 to 200 GeV. For the sixth, the antiproton, the rapid decrease in size previously observed below 50 GeV has dramatically slowed and the apparent size becomes essentially constant between 150 and 200 GeV.

The similarities and the intercomparisons of the behaviour of the cross sections with the six probing particle beams indicate that a new simplicity of nature may be revealing itself at very high energies; a situation which has been predicted by some physicists. Since the proton and neutron are the basic blocks of all atomic nuclei, these experiments are a significant advance toward an understanding of the constitution of matter.

The phenomenon of cross sections rising with energy was first suggested in 1971 by scientists working with a beam of positively charged K-mesons at energies up to 55 GeV at the U.S.S.R. Serpukhov Accelerator.

In 1973, scientists working at the European Center for Nuclear Research (CERN) near Geneva, Switzerland announced an increase in proton-proton total cross sections. Although the CERN scientists were limited to the study of proton-proton collisions by the nature of the interacting Storage Ring Accelerator, in which two oppositely directed beams of protons collide with each other, they were able to reach a record equivalent energy of 2,000 GeV.

The new measurements just announced were made with the world's; largest accelerator recently dedicated at Fermilab in which 300 GeV protons strike a stationary nuclear target. Beams of many types of subnuclear particles with energies up to 300 GeV emerge from this target. The six varieties of particle beams which were used included protons, antiprotons, positively and negatively charged pi-mesons and positively and negatively charged K-mesons.

The experiment was carried out by a collaboration between teams of physicists from the Brookhaven National Laboratory (BNL) in Upton, N.Y., the Fermilab in Batavia, Illinois, and The Rockefeller University in New York City.

Details of today's announcement were disclosed at a press conference at the American Institute of Physics. Participating in the conference were: Dr. Thaddeus F. Kycia of BNL; Dr. Winslow F. Baker of Fermilab; and Professor Rodney L. Cool of the Rockefeller University.

Other menbers of the teams and authors of the paper were as follows:

BNL – -Alan S. Carroll, I-Hung Chiang, Kelvin K. Li, Peter O. Mazur, Paul M. Mockett, now at the University of Washington, Seattle, and David C.Rahm.

Fermilab- - David P. Eartly, Georgio Giacomelli, from the Institute of Physics at the University of Padova, Italy, Peter F. M. Koehler, Klaus P. Pretzel, now at the Max Planck Institute of Physics and Astrophysics in Munich, Germany, Roy Rubinstein, and Alan A. Wehmann.

The Rockefeller University - - Orrin D. Fackler

The principal findings announced today are:

1. The increase of the size of a proton with increasing energy appears to be a general and systematic property of strongly interacting nuclear forces. Five of the particles employed were the pi-meson, plus and minus, the K-meson, plus and minus, and the proton itself. The cross sections of protons measured with each of these probes increased with energy between 50 and 200 GeV. The cross section measured with the sixth probe the antiproton, ceased to fall and became constant between 150 and 200 GeV.

    The experiment will be continued up to 400 GeV. When the higher energies are employed in this same experiment, it is possible that all of the cross sections will rise. If this behavior is a universal phenomenon, it will give strong clues as to the fundamental character of the strong nuclear interactions and may assist in reaching a general theory of the strong nuclear forces which has been sought for many years.

2. All of the particle-proton and antiparticle-proton cross section pairs uniformly approach each other approximately inversely proportional to the square root of their energy. The theorem that the difference between a particle cross section and that of its antiparticle on the same target should approach zero at very high energies was enunciated by the Russian theorist, Isaak Ya. Pomeranchuk in 1958.

3. Since it is not possible to have a target of pure neutrons, the other basic ingredient of the atomic nucleus, the deuteron—which is composed of one proton and one neutron—was used as a target. By comparing the proton cross sections and the deuteron cross sections, the neutron cross sections were deduced. For each probe particle, the neutron cross section is very nearly equal to the proton cross section at these ultra high energies.

4. The nuclear forces continue to be charge symmetric. The cross section for a charged pion on a proton is equal to that of the oppositely charged pion on a neutron.

5. This experiment provides the "systematics" of the behavior of an entire class of interactions. The proton and neutron appear to have cross sections nearly equal to each other for all of the probes. The differences between particle and antiparticle pairs seems to be disappearing at extremely high energies. Even the difference in cross section between particles which do not have the same quantum number describing the characteristic of "strangeness" seems to be disappearing.
   [...]

The measurements performed later in the momentum range 200-340 GeV/c confirmed the above statements and proved that also the antiproton-proton total cross section was rising with energy, see Fig. 11. The differences of (antiparticle-proton) – (particle-proton) cross sections are shown in Fig. 12.

The rising of the total cross sections at high energies was a surprise to most physicists. I remember the heated discussions concerning the high energy behaviour of the cross sections; one of these discussions was made at a coffee table by a group of experimentalists and theoreticans: most of the experimentalists favoured a constant cross section, while most theoreticians favoured cross sections becoming smaller. Then arrived Giuseppe Cocconi (one of the driving forces of high energy total and elastic cross section measurements): He listened carefully, then said, "it is all nonsense: I bet you a coffee that total cross sections will rise". This was the first time I, and most experimental colleagues, had heard of this possibility, which looked somewhat ridiculous. Therefore we accepted the bet and a few years later we had to pay it.

The study of total cross sections requires first a study of the beam qualities and of their fluxes. This in turns provides interesting information on the production cross sections of the six long lived charged hadrons, see for instance Fig.13 [22].

Besides the liquid hydrogen, liquid deuterium and dummy targets, one had always available a number of targets of different materials (Li, C, Al, Cu, Sn and Pb). Thus one had the possibility of measuring the absorption cross sections in nuclei, as shown in Fig. 14 [10, 13, 19].

## 4. The "continuation"

As already stated, the participation in the Brookhaven and Fermilab total cross section experiments created a strong group of friends interested in this line of research. Some of these collaborators made further measurements at different accelerators [18-22].

The logical continuation of the total cross section measurements performed at the fixed target BNL, Serpukhov, and Fermilab accelerators and at the CERN ISR was to measure the total antiproton-proton cross section at the CERN [23] and Fermilab $\bar{p}p$ colliders [24], up to 1.8 TeV c.m. energy. Several members of the previous collaborations measured the antiproton-proton total and elastic cross sections at Fermilab [24]. Clearly, at a collider, one needs a layout considerably different from that used for the transmission measurements performed at fixed target accelerators.

The Fermilab collider results established that the total PN cross section for antiproton-proton interactions keeps increasing with increasing c.m. energies, at least up to 1.8 TeV. Cosmic ray measurements indicate that the total cross sections increase even at higher energies, see Fig. 11b.

The next steps will be measurements at RHIC at BNL and then at the Large Hadron Collider (LHC, the proton-proton collider which is being built at CERN). Maybe some of the collaborators will participate in those experiments

## Conclusions

The series of total cross section measurements performed at BNL in the resonance region lead to the discovery of a large number of peaks and structures, most of which correspond to hadronic resonances.

Since the early Serpukhov data, then the CERN data and the Fermilab data we know that all total hadron-hadron cross sections increase with energy; this is also confirmed, even if with low precision, by the highest energy cosmic ray data.

From a theoretical point of view there is not a unique interpretation for this rise, though in many QCD inspired models it may be connected with the increase of the number of minijets and thus to semi-hard gluon interactions.

Most of the high energy elastic and total cross section data have been usually interpreted in terms of Regge Poles, and thus in terms of Pomeron exchange. Even if the Pomeron was introduced long time ago we do not have a consensus on its exact definition and on its detailed substructure. Some authors view it as a "gluon ladder".

Future experiments on hadron-hadron total cross sections remain for the moment centered on the Fermilab Collider, (for $\bar{p}p$)where new measurements will be made in the near future. The experimental future after year 2005 will rely mainly on the LHC proton-proton Collider at CERN. Large area cosmic ray experiments may be able to improve the data in the ultra high energy region.

In any case the collaborations headed by Ted Kycia and Rod Cool produced important results which will remain known in the history of particle physics. "Ted's expertise was in the design, planning, and execution of particle physics experiments, and he had an impressive record of obtaining correct and accurate results. We and many of our colleauges learned much through working with him".

I am greatful to all the collaborators in the various total cross section measurements. I am also greatful to their families for the nice atmosphere during and after work. I thank Ms. Luisa De Angelis for typing the manuscript and Roberto Giacomelli for technical support.